\documentclass[prl,twocolumn,english,showpacs,floatfix,superscriptaddress]{revtex4}

\usepackage{graphics}
\usepackage{graphicx}
\usepackage{epsfig}
\usepackage{amssymb}
\usepackage{amsmath}
\def\beq{\begin{equation}}
\def\eeq{\end{equation}}

\begin{document}

\newcommand{\ket}[1]{|{#1}\rangle}
\newcommand{\bra}[1]{\langle{#1}|}
\newcommand{\braket}[1]{\langle{#1}\rangle}
\newcommand{\ad}{a^\dagger}
\newcommand{\e}{\ensuremath{\mathrm{e}}}
\newcommand{\norm}[1]{\ensuremath{| #1 |}}
\newcommand{\aver}[1]{\ensuremath{\big<#1 \big>}}
\renewcommand{\Im}{\operatorname{Im}}
\newcommand{\etal}{\textit{et al.~}}
\newcommand{\p}{\ensuremath{\partial}}
\newcommand{\br}{\mathbf{r}}

\title{Trapping and cooling fermionic atoms into the Mott and N\'eel states}

\author{Lorenzo De Leo}
\affiliation{Centre de Physique Th\'eorique, Ecole Polytechnique, CNRS,
91128 Palaiseau, France}
\author{Corinna Kollath}
\affiliation{Centre de Physique Th\'eorique, Ecole Polytechnique, CNRS,
91128 Palaiseau, France}
\author{Antoine Georges}
\affiliation{Centre de Physique Th\'eorique, Ecole Polytechnique, CNRS,
91128 Palaiseau, France}
\author{Michel Ferrero}
\affiliation{Centre de Physique Th\'eorique, Ecole Polytechnique, CNRS,
91128 Palaiseau, France}
\author{Olivier Parcollet}
\affiliation{Institut de Physique Th{\'e}orique, CEA/DSM/IPhT-CNRS/URA 2306 CEA-Saclay,
F-91191 Gif-sur-Yvette, France}

\begin{abstract}
We perform a theoretical study of a fermionic gas with two hyperfine states
confined to an optical lattice. We derive a generic state diagram as a function
of interaction strength, particle number, and confining potential. We discuss
the central density, the double occupancy and their derivatives
as probes for the Mott state, connecting our findings to the recent experiment of J\"ordens
\etal~\cite{JoerdensEsslinger2008}. Using entropic arguments we compare two
different strategies to reach the antiferromagnetic state in the presence of
a trapping potential.
\end{abstract}

\maketitle

Remarkable experimental progress has been made in handling and controlling
quantum degenerate atomic gases. The confinement of these gases within optical
lattices, and the use of Feshbach resonances allow for the investigation of
strongly correlated quantum phases~\cite{BlochZwerger2008}.
An outstanding example is the experimental study of the superfluid to Mott
insulator transition in bosonic gases~\cite{GreinerBloch2002}. A Mott
insulating state is characterized by the suppression of density fluctuations.
Very recently, the observation of an incompressible Mott state has been
reported in a fermionic gas using two hyperfine states of
$^{40}K$~\cite{JoerdensEsslinger2008}.
A future major step will be the observation of the antiferromagnetic ordering
of the two hyperfine components in the lattice.
However, whereas an incompressible Mott state is reached when the temperature
is low enough as compared to the Mott gap, stabilizing a magnetically ordered state
requires cooling down to a lower temperature scale set by the superexchange inter-site
coupling.

In this article, we address theoretically several issues which have immediate relevance
to those experimental efforts.
We establish a state diagram of cold fermions subject to an optical lattice
and a trapping potential, as a function of the number of atoms and coupling strength.
We discuss how to detect different states, by considering observables such as
the fraction of doubly occupied sites, the local occupancy and the compressibility
at the center of the trap.
Finally, we outline two possible strategies for adiabatically cooling the system into
a long-range ordered antiferromagnet.

A quantum degenerate mixture of two hyperfine states of fermionic atoms can be
described by a Hubbard-type
Hamiltonian~\cite{JakschZoller1998,HofstetterLukin2002}:
\begin{eqnarray}
   \label{eq:h}\nonumber
   H&&= -J \sum_{\langle j,j'\rangle,\sigma} \left(c_{j,\sigma}^\dagger
   c^{\phantom{\dagger}}_{j',\sigma}+h.c.\right)
   + U \sum _{j} \hat{n}_{j,\uparrow} \hat{n}_{j,\downarrow} + \,\\
   &&  -\, \mu_0 \sum_{j,\sigma}\,\hat{n}_{j,\sigma}
   +\, V_t \sum_{j,\sigma}(r_j/d)^2\,\hat{n}_{j,\sigma}
\end{eqnarray}
Here, $\sigma=\uparrow , \downarrow$ labels the two hyperfine states,
$j$
the sites of a three-dimensional (3D) cubic lattice, and $\langle
j,j'\rangle$ neighboring sites. $r_j$ is the distance between $j$ and the
center, $d$ the lattice spacing, and $\mu_0$ the chemical potential.
We consider a mixture with equal number
$N/2$ of $\uparrow$ and $\downarrow$ atoms.
Experimentally the interaction parameter $U$ and the hopping parameter $J$ can
be tuned independently changing the optical lattice height and using a Feshbach resonance.
Note that the description by a one-band model is only
valid for large enough lattice depth, small enough scattering length, and low
temperatures, i.e. situations in which the higher bands and more complicated
interaction or hopping terms can be
neglected~\cite{JakschZoller1998,WernerHassan2005}.

\begin{figure}[tb]
  \centerline{
    \includegraphics[width=0.8\linewidth]{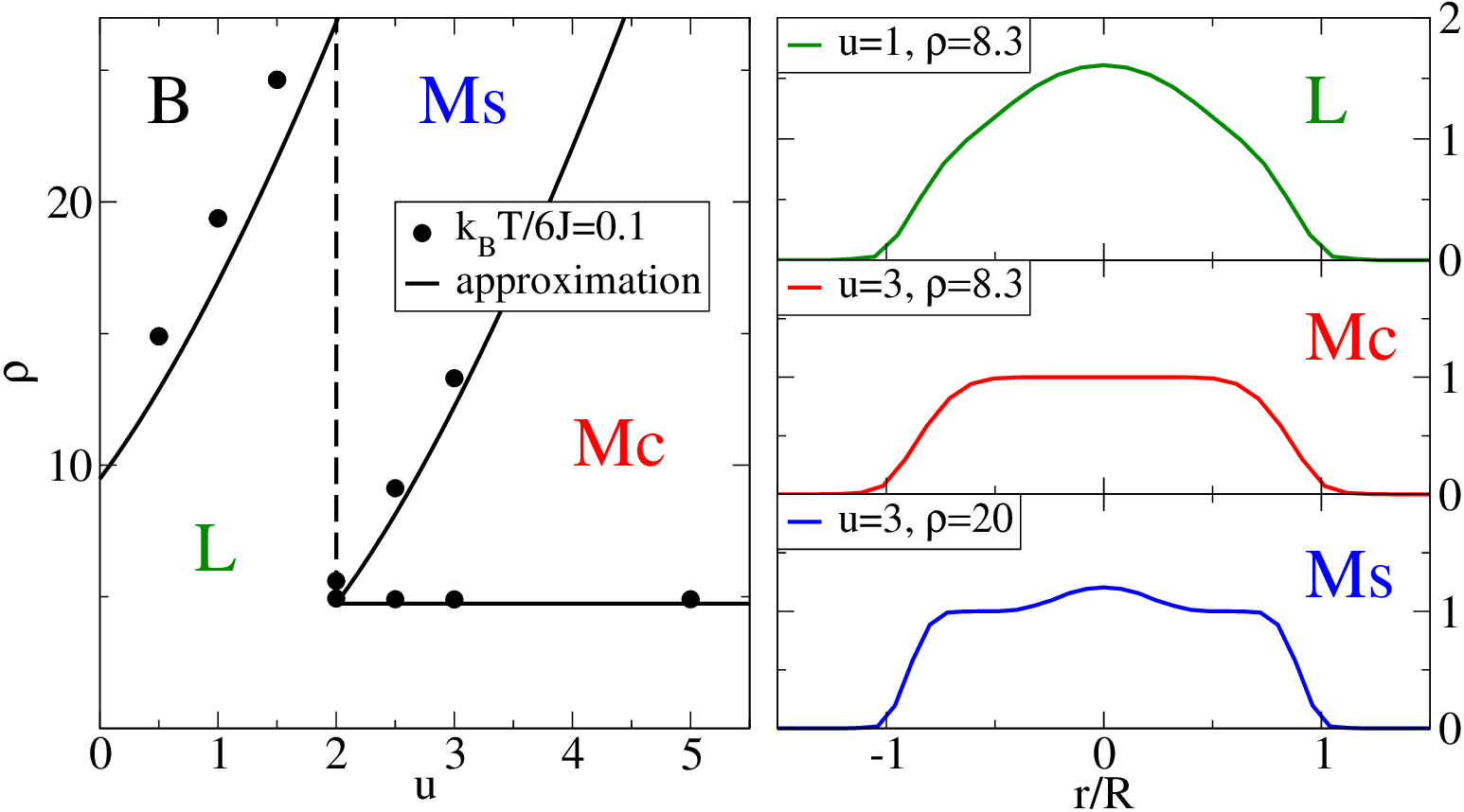}
  }
  \caption{(Color online) Left panel: State diagram for fermionic atoms
  as a function of interaction
  strength $u\equiv U/(6J)$ and scaled particle number
  $\rho\equiv N (V_t/6J)^{3/2}$.
  Symbols show results of DMFT calculations.
  The boundaries of the region (Mc) with a Mott-insulator in the center of the trap
  are defined from the constraint ~$\norm{n_0-1}<0.005$ on the central density.
  Lines are an analytical continuum approximation at $T=0$.
  Right panel: typical density profiles for regions (L), (Mc) and
  (Ms), with the radius $R$ set by $n(R)=0.1$.
  }
  \label{fig:state}
\end{figure}

One important effect of the confining potential is to allow for the spatial
coexistence of different quantum
states~\cite{BatrouniTroyer2002,KollathZwerger2003}.
For a 1D tube of fermionic atoms, a Mott-insulating phase
next to a liquid phase was pointed out \cite{RigolScalettar2003,LiuHu2005}.
More recently, the stability of antiferromagnetically ordered regions has been
investigated in a 2D system \cite{SnoekHofstetter2008}.
Helmes \etal \cite{HelmesRosch2008} studied the coexistence of the liquid
and insulating phases, in 3D, calculating in particular the density and
momentum profiles and the local spectral function.
In Fig.~\ref{fig:state}, we display a state diagram which summarizes the
different possible shapes of the density profiles as a function of coupling
strength and atom number at the temperature $k_BT/(6J)= 0.1$.
All calculations in this paper were obtained
using a local density approximation (LDA), in combination with
dynamical mean field theory (DMFT) \cite{GeorgesRozenberg1996,WernerMillis2006}
to simulate the homogeneous system in the paramagnetic state.
%

In order to compare different trapping energies and particle numbers, we use the
scaled number of atoms $\rho=N(V_t/6J)^{3/2}$ (see
Ref. \cite{RigolScalettar2003} for 1D).
Within LDA and the continuum limit, $\rho$ entirely determines the shape of the
density profile at a fixed interaction strength and temperature,
allowing for a straightforward comparison to experimental setups.
The state diagram in Fig.~\ref{fig:state} displays four distinct regimes, labeled by
(L), (Mc), (Ms) and (B).
(L): For small interaction or small density, the whole system is in a (Fermi-)
liquid phase. There, the density profile depends strongly on the trapping
potential. (Mc): Above a critical interaction strength $u\equiv U/(6J) \gtrsim
2$ and a characteristic density $\rho\gtrsim 5$
a Mott central plateau with $n\approx 1$ forms.  A
liquid phase surrounds this plateau. 
(Ms): When the particle number is further increased
the pressure exerted by the trapping potential overwhelms the incompressibility
of the Mott state. This leads to a state with a central occupancy larger than one
particle per site surrounded by a Mott insulating shell. (B): For even larger
particle numbers a band-insulating plateau with $n=2$ forms, surrounded by liquid and possibly Mott-insulating shells.

To detect the occuring quantum phases we investigated mainly two observables: the occupancy at the center of
the trap $n_0$ and the fraction of atoms residing on doubly occupied sites
$D=\frac{2}{N}\sum_j \langle n_{j\uparrow}n_{j\downarrow} \rangle$.
$D$ was measured by molecule formation in the experiments of
Ref.~\cite{JoerdensEsslinger2008}. The experimental observation of the density 
in a small central region is more involved. However, a lot of effort is currently devoted to
the development of probes with good spatial resolution.
To make contact with current experiments, we assume the slow experimental
loading process \cite{JoerdensEsslinger2008} into the optical lattice to be
adiabatic, and perform calculations at a fixed
value of the entropy per particle, $s$~\footnote{The entropy
of the homogeneous system is obtained by integrating a fit to the DMFT results
for the energy. The high temperature regime is approximated using a high temperature expansion.}. The latter is related to the
initial temperature of the non-interacting fermion gas in the
absence of the lattice by $T_i/T_F = s/\pi^2 +O[(T_i/T_F)^3] $.

\begin{figure} [tb]
  \centerline{
    \includegraphics[width=0.85\linewidth]{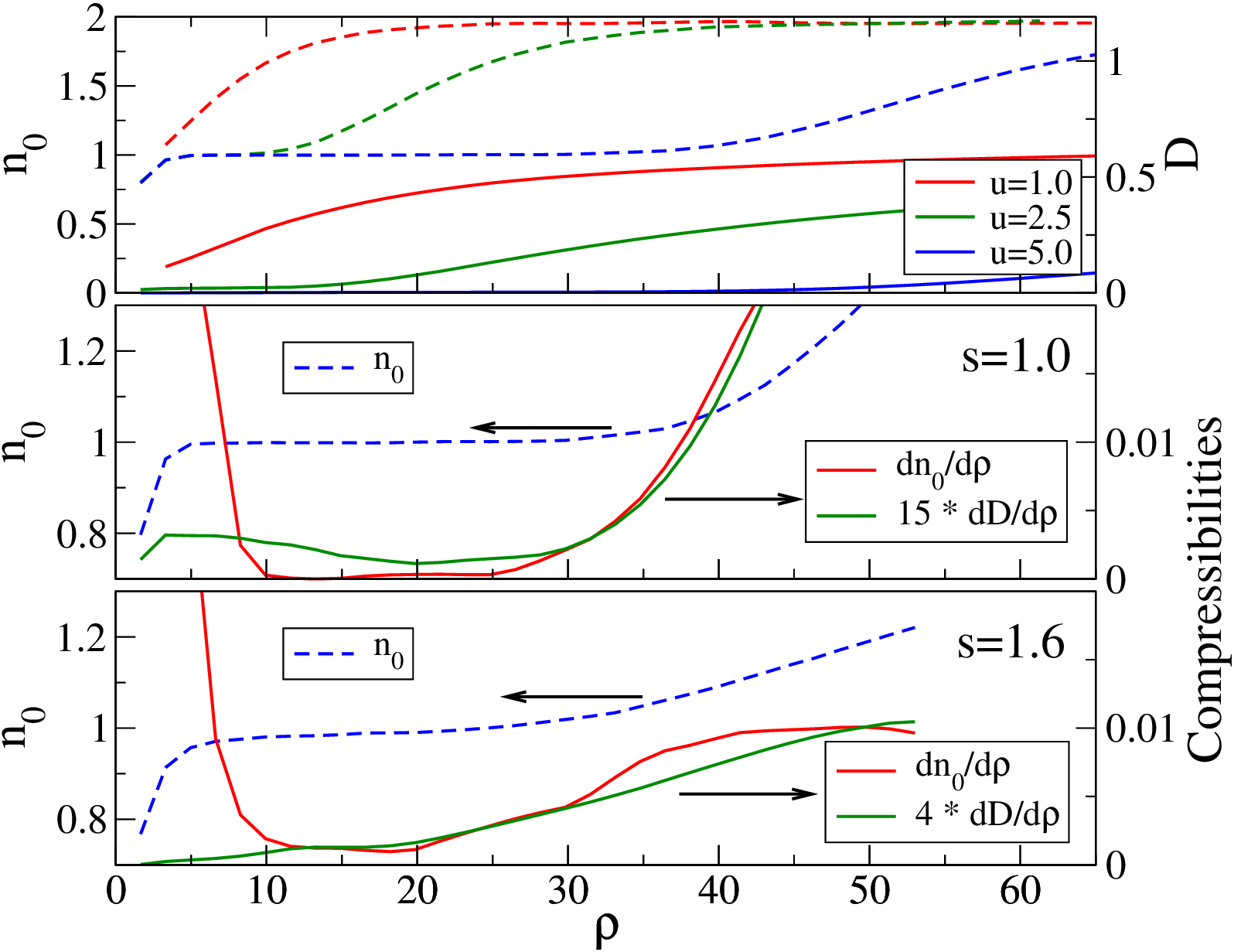}
  }
  \caption{(Color online) First panel: Occupancy $n_0$ at the center of the trap
  and fraction $D$ of atoms residing on doubly occupied sites
  versus the scaled particle number $\rho$ at constant initial temperature $T_i/T_F= s/\pi^2=1.0/\pi^2$.
  The central occupancy $n_0$ and the ``compressibilities''
$\partial n_0/\partial\rho$ and $\partial D/\partial\rho$, for $u=5$, at two different
initial temperatures $T_i/T_F =1.0/\pi^2,1.6/\pi^2$ are shown in the second and the third panel, respectively.
  }
  \label{fig:DvN}
\end{figure}
The first panel of Fig.~\ref{fig:DvN} displays $n_0$ and $D$ as a function of
particle number at a fixed initial temperature $T_i/T_F=1.0/\pi^2$.
As we detail later on, a plateau in $n_0$ {\it vs.} $\rho$ provides a good indicator
of the formation of an incompressible Mott insulating region in the center of the trap.
The three sets in Fig.~\ref{fig:DvN} (first panel) correspond to different
regimes of coupling.
For weak interactions ($u\lesssim 1.5$), no plateau is present:
$n_0$ smoothly increases and saturates at $n_0\approx 2$ corresponding to the
regimes (L) and (B). At intermediate coupling
$u\gtrsim 2.5$, a Mott plateau starts forming ($\rho \approx 5$), corresponding to regime (Mc).
$n_0$ eventually deviates from unity as $\rho$ is further
increased into regime (Ms).
With increasing coupling (cf.~$u \gtrsim 5$), the central plateau
broadens.
From Fig.~\ref{fig:DvN} (first panel), it is also clear that there is a qualitative
correlation between the existence of a Mott central plateau and the behavior of
$D$. At weak coupling, $D$ just increases smoothly as $\rho$ is increased.
At stronger coupling, $D$ remains very small for the low values of $\rho$, and starts
raising approximately at the transition from the central incompressible regime (Mc)
into regime (Ms).

We now make this correlation between $D$ and $n_0$ more quantitative
ploting in the lower panels of Fig.~\ref{fig:DvN} the derivatives
$\partial n_0/\partial\rho$ and $\partial D/\partial\rho$, for $u=5$, at two different
initial temperatures $T_i/T_F =1.0/\pi^2,1.6/\pi^2$.
In the LDA approximation, $\partial n_0/\partial\rho$
is directly proportional to the local compressibility at the center of the trap
$\kappa_0=\partial n_0/\partial\mu_0$, through:
$\rho\,\partial n_0/\partial\rho = \kappa_0/\langle\kappa\rangle$. In this expression,
$\langle\kappa\rangle$ is an average of the compressibility over the whole trap.
The average $\langle\kappa\rangle$ is a non-singular quantity since it always has sizeable contributions from the
liquid wings of the density profile. Hence, $\partial n_0/\partial\rho$ is
a direct measure of the local incompressibility and its vanishing signals the Mott state. 
Indeed, it is seen from Fig.~\ref{fig:DvN} (second panel) that, for the
lowest temperature studied $T_i/T_F =1.0/\pi^2$, this derivative
is very small ($\lesssim 10^{-4}$) inside the (Mc) regime, and increases dramatically
as one departs from a Mott insulator in the trap center. For the higher temperature
$T_i/T_F =1.6/\pi^2$ (Fig.~\ref{fig:DvN}, third panel), $\partial n_0/\partial\rho$ is only of order
$10^{-3}$ in the rather narrow (Mc) regime, and increases less steeply as $\rho$
is increased.
In both cases, it is apparent from Fig.~\ref{fig:DvN} that, remarkably, the dependence of
$\partial D/\partial\rho$ on $\rho$ is very similar to that of the
`compressibility' $\partial n_0/\partial\rho$, for values of $\rho$ in the
(Mc) and (Ms) regimes.
The reason for this observation is that, despite being a global quantity over the
whole trap, the double occupancy fraction remains very small in the boundary regions due to the low density.
Therefore it is dominated by the central regions with occupancies $n\gtrsim 1$.
Hence $\partial D/\partial\rho$ is small in regime (Mc)
and increases sharply when the incompressibility at trap center is broken.
At smaller values of $\rho$, these quantities are obviously very different: $D$ is very
small due to the low filling whereas $\partial n_0/\partial\rho$ is finite due
to the liquid character of the state.
Thus, we conclude that $D$ and $\partial D/\partial\rho$
are rather good indicators of when the Mott state
ceases to exist at trap center.
This validates the use of these quantities as a
diagnostic for a Mott insulator region, as done in 
Ref. \cite{JoerdensEsslinger2008}.
A closely related observation was made by Scarola \etal \cite{Scarola08}, 
who further supported it by showing that the compressibility 
$\partial n/\partial\mu$ is essentially identical to 
$\partial (D/N) /\partial\mu$ in the homogeneous system for $n\gtrsim 1$.

Comparing directly to the experimental results shown in Fig.~2 of
Ref.~\cite{JoerdensEsslinger2008}, we find good agreement
between the upturn of $D(\rho)$ in
the experimental data for $U/6J=4.8$ approximately at $\rho=20$ 
and our theoretical results for $U/6J=5$ (cf. Fig.~\ref{fig:DvN}). 
However, in this regime of coupling, the estimated lowest temperature in
the experiments ($T_i/T_F\simeq 0.15$), while being at the onset of the incompressible Mott regime,
still implies a sizeable contribution of thermal excitations within the Mott gap
(Fig.~\ref{fig:DvN} third panel). In contrast, at lower temperature (second panel) a much flatter plateau and
lower value of the compressibility are obtained. Obviously,
a Mott plateau forms at even higher temperature for stronger interactions,
but then equilibration times may become an issue.

\begin{figure} [tb]
  \centerline{
  \includegraphics[width=0.8\linewidth]{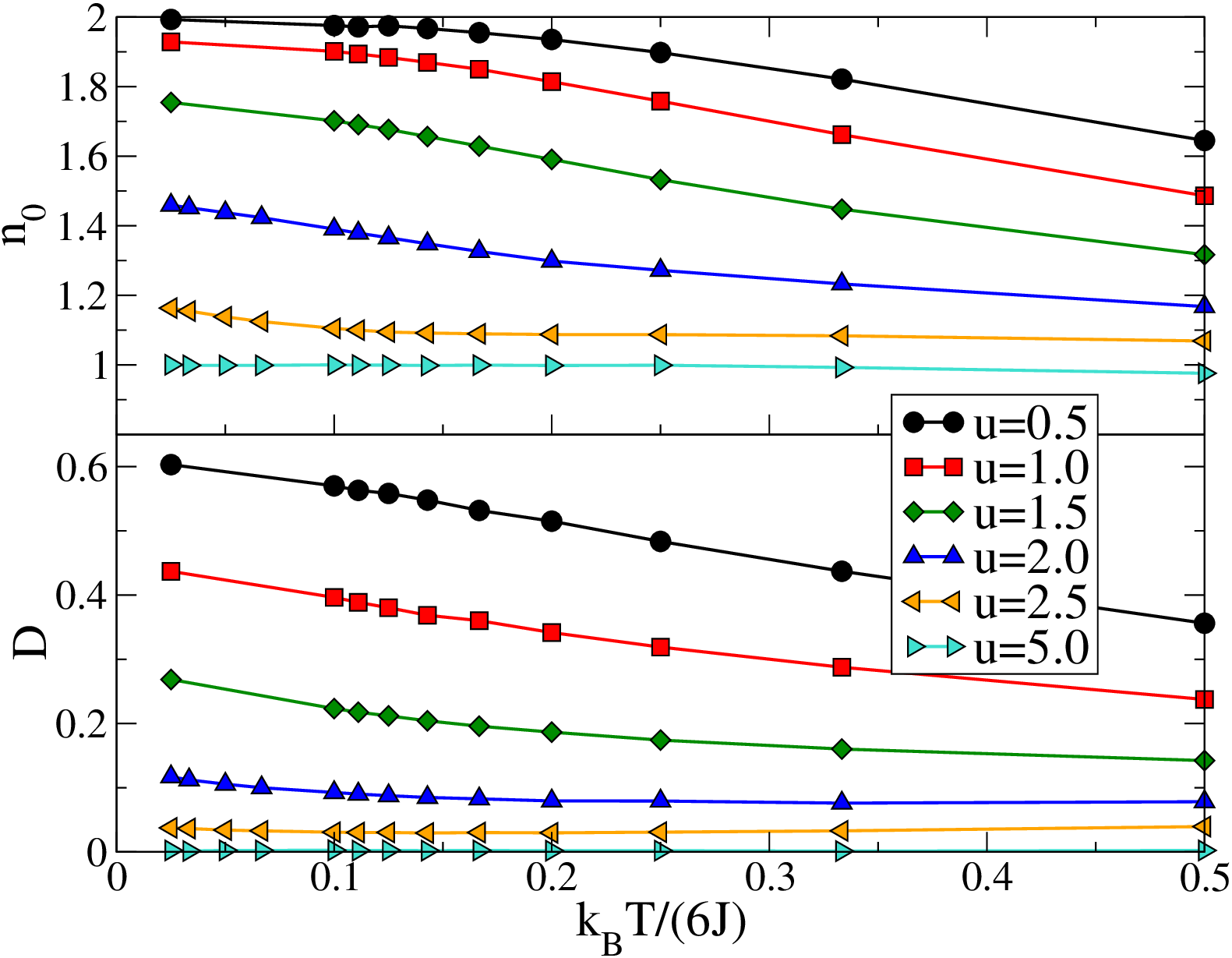}
  }
  \caption{(Color online) Upper panel: Central occupancy $n_0$ versus
  temperature for different interaction strengths. Lower panel: double occupancy
  $D$ versus temperature. ($\rho\approx 13.3$)
  }
  \label{fig:DvT}
\end{figure}

A crucial point in experiments is to estimate the temperature of the
interacting fermionic system in the presence of an optical lattice.
For a confined non-interacting gas the double
occupancy fraction has been used experimentally \cite{StoeferleEsslinger2006} 
as a thermometer since $D(T)$ {\it decreases} as the temperature is increased 
\cite{Koehl2006,KatzgraberTroyer2006}.
We find that the decrease of $D$ with temperature persists throughout the
weak and intermediate interaction regimes ($u\lesssim 1.5$, 
Fig.~\ref{fig:DvT}, lower panel).
%
%
As in the non-interacting gas, the transfer of atoms from the trap center
(cf.~Fig.~\ref{fig:DvT}, upper panel) towards the trap
boundaries, due to thermal fluctuations, causes this
effect.
It leads to a decrease of $D$ which overwhelms the increase
found in the high-$T$ or large-$U$ regime for the homogeneous system.
Also, the Pomeranchuk effect \cite{WernerHassan2005,
DareTremblay2007} at low-$T$ and intermediate $U$
reinforces this trend.
In contrast, for stronger interactions, $D$ hardly changes with temperature
($u \lesssim 2.5$, Fig.~\ref{fig:DvT}) and ceases to be useful for thermometry
because the density profile becomes almost frozen as the incompressible Mott 
state is approached.

One major step after observing the Mott regime will be the realization of the
antiferromagnetically (AF) ordered phase, which
occurs at a still lower temperature $T_N$ (N\'eel temperature). 
We now discuss two possible strategies
to reach this phase.
(a) A first possibility is to work at intermediate coupling ($u\simeq 2.5$), 
for which $T_N$ is the largest.
(For possible ways of increasing $T_N$ further, see \cite{MathyHuse2008}).
The drawback of this strategy is that a sizeable Mott insulating region can
only be stabilized within a narrow range of particle numbers
(cf.~Fig.~\ref{fig:state}).
(b) The second possibility is to work at stronger coupling (e.g. $u\simeq 5$)
so that a Mott region can be stabilized over a much wider range of
particle numbers (Fig.~\ref{fig:state}).
Since $T_N\propto J^2/U$ at large $U$, this requires
cooling the gas down to very low temperatures.
However, assuming an adiabatic evolution,
the entropy per particle at which the ordering takes place, $s(T_N)$,
is the key quantity to focus on, rather than the absolute temperature $T_N$.
In Ref.~\cite{WernerHassan2005}, it was pointed out that,
for the homogeneous half-filled system, $s_h(T_N)$
reaches a {\it finite} value in the strong-coupling limit, even though $T_N$
is very small in this limit.
Here, we compare the initial temperatures $T_i/T_F$ needed to reach the AF
ordered state for the two strategies (a) and (b), in the presence of the
trapping potential \footnote{To reach the AF state, the correlation length 
needs to exceed the spatial extension of the central Mott region. However, for a
large number of atoms this leads to negligible corrections as compared to a
criterion for the bulk.}.

Once an incompressible Mott region is formed in the center of the trap, the
entropy is the sum of two contributions: one from the liquid wings of
the density profile, and one from the central Mott region.
The entropy of the Fermi-liquid fraction is linearly increasing with
temperature, below some characteristic coherence temperature (which is larger
at low density and smaller for densities close to one particle per site).
By contrast, the entropy in the Mott state
remains almost constant upon cooling.

\begin{figure} [tb]
  \centerline{
  \includegraphics[width=0.8\linewidth]{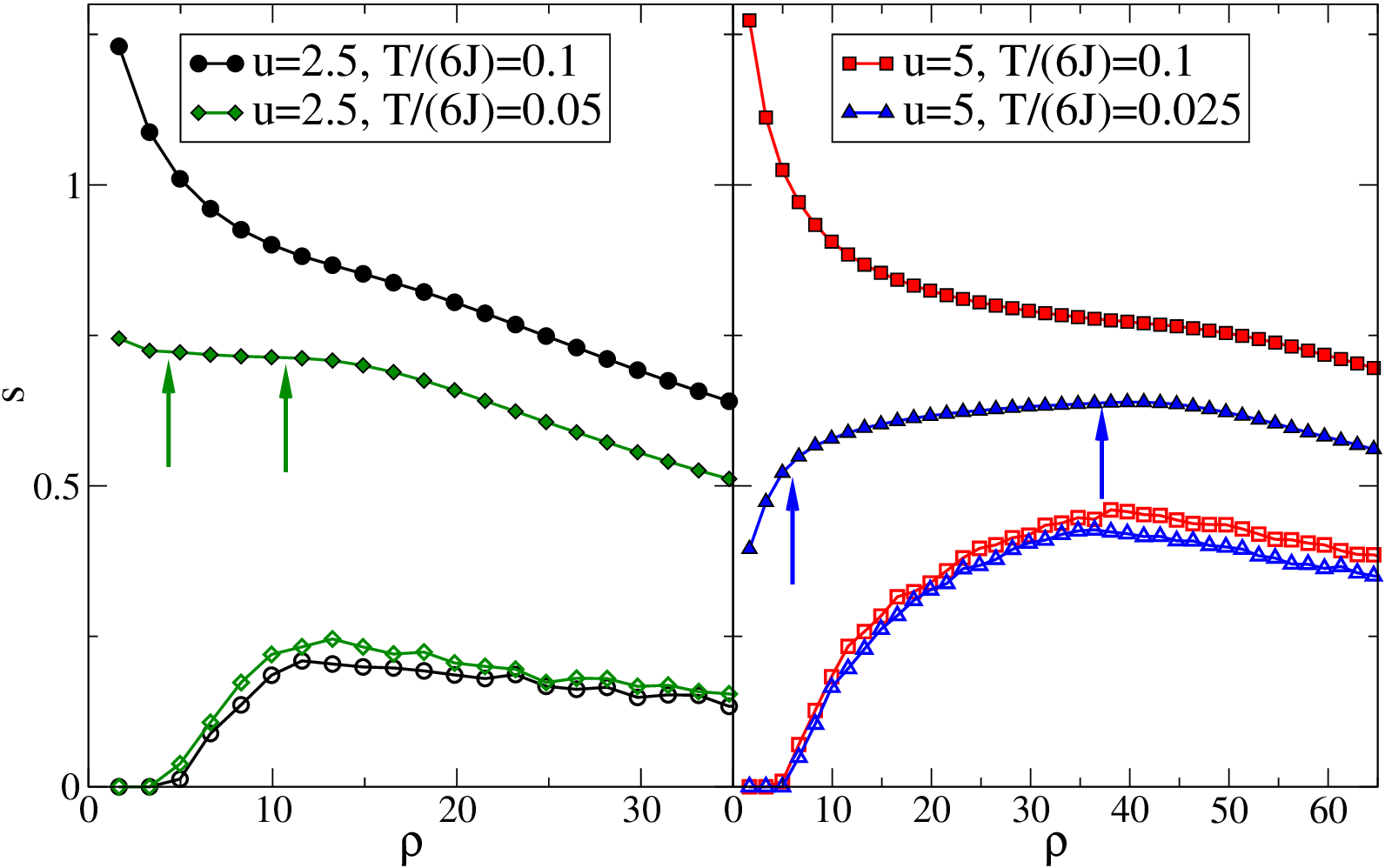}
  }
  \caption{(Color online) Entropy per particle $s$ of the total system (filled
  symbols) and the Mott region (open symbols) versus characteristic density $\rho$.
  The lowest $T$ is chosen slightly above the N\'eel temperature and the arrows
  delimit the density range where a central Mott plateau is formed.
  }
  \label{fig4}
\end{figure}

At intermediate coupling (a), the contribution of the liquid wings dominates
over that of the central Mott region in the whole temperature range above
$T_N$ (Fig.~\ref{fig4}, left panel). This is because the fraction
of atoms in the central region is rather small ($\sim30\%$) and 
$T_N$ in this regime is high enough such that part of the wings have not
yet reached Fermi-liquid coherence. Hence, strategy (a) can benefit from the
high entropic contributions of the liquid wings. The lowest temperature
displayed in Fig.~\ref{fig4} is close to $T_N$ at $u=2.5$.
The entropy per particle which must be reached is of order
$s\simeq 0.67$ (taking $\rho\simeq 5-10$ where a Mott central region is
stabilized). This corresponds to an initial temperature $T_i/T_F\simeq 0.07$.

For stronger interaction (b), the density profile is dominated by a wide
Mott-insulating region, for a large range of particle numbers.  The entropy of
this region gives a major contribution to the total entropy already at
relatively large temperatures ($\sim 50\%$ at $k_BT/6J=0.1$ as shown in
Fig.~\ref{fig4}, right panel) and it
does not change much upon cooling. In contrast, the contribution of the liquid
wings decreases strongly and eventually becomes very small (30\% at
$k_BT/6J=0.025$). In this regime, DMFT yields a value of the
entropy per particle of order $0.6$ near $T_N$, comparable to
strategy (a) above.  However, DMFT overestimates the entropy
of the Mott state. We propose a lower bound on
the entropy at $T_N$ by neglecting the contribution of the
liquid wings, and considering only that of the Mott region. This leads
to the estimate $s= s_h(T_N)\,N_c/N$. Using a typical value $N_c/N=2/3$ for the
fraction of atoms in the central Mott region, and the entropy per particle
of the homogeneous Mott state
at $T_N$, $s_h(T_N)\approx \ln 2/2$ (as estimated from
fluctuation effects in the Heisenberg model~\cite{WernerHassan2005}, see
also~\cite{KoetsierStoof2008}), we obtain $s\simeq 0.2$.  This corresponds to
an initial cooling down to $T_i/T_F\simeq 0.02$ to reach AF ordering.
We view this estimate as a pessimistic lower bound.

To summarize we established a state diagram of cold fermions in an optical lattice
and a trapping potential. We further showed how to detect the Mott-insulating state using
the central density, the fraction of doubly occupied sites and their derivatives.
The latter were shown to be related to the compressibility at the center of the trap.
We found that the double occupancy is well suited for thermometry in the lattice at weak and
intermediate coupling, but not at strong coupling.
Finally, we compared two strategies for adiabatically cooling into
a long-range ordered antiferromagnet.
The intermediate coupling route (a) is somewhat more favorable
from the cooling point of view. By contrast the stronger coupling route (b) is more
favorable in terms of the number of atoms and extension of the Mott plateau,
while the corresponding ordering temperatures may still be reachable.

{\em Note added.}---After the completion of this work also Schneider \etal (arXiv:0809.1464)
reported the observation of an incompressible Mott state.


We acknowledge fruitful discussions with E.~Demler, M.~K\"ohl, W.~Ketterle,
C.~Salomon, M.~Zwierlein, the group of T.~Esslinger and I.~Bloch. 
Support was provided by
the `Triangle de la Physique', the DARPA-OLE program, ICAM and the Agence
Nationale de la Recherche under contracts GASCOR and FABIOLA.


\begin{thebibliography}{20}
\expandafter\ifx\csname natexlab\endcsname\relax\def\natexlab#1{#1}\fi
\expandafter\ifx\csname bibnamefont\endcsname\relax
  \def\bibnamefont#1{#1}\fi
\expandafter\ifx\csname bibfnamefont\endcsname\relax
  \def\bibfnamefont#1{#1}\fi
\expandafter\ifx\csname citenamefont\endcsname\relax
  \def\citenamefont#1{#1}\fi
\expandafter\ifx\csname url\endcsname\relax
  \def\url#1{\texttt{#1}}\fi
\expandafter\ifx\csname urlprefix\endcsname\relax\def\urlprefix{URL }\fi
\providecommand{\bibinfo}[2]{#2}
\providecommand{\eprint}[2][]{\url{#2}}



\bibitem[{\citenamefont{J\"ordens et~al.}(2008)\citenamefont{J\"ordens,
  Strohmaier, G\"unter, Moritz, and Esslinger}}]{JoerdensEsslinger2008}
\bibinfo{author}{\bibfnamefont{R.}~\bibnamefont{J\"ordens \etal}},
  \bibinfo{journal}{Nature} \textbf{\bibinfo{volume}{455}},
  \bibinfo{pages}{204} (\bibinfo{year}{2008}).

\bibitem[{\citenamefont{Bloch et~al.}(2007)\citenamefont{Bloch, Dalibard, and
  Zwerger}}]{BlochZwerger2008}
\bibinfo{author}{\bibfnamefont{I.}~\bibnamefont{Bloch}},
  \bibinfo{author}{\bibfnamefont{J.}~\bibnamefont{Dalibard}}, \bibnamefont{and}
  \bibinfo{author}{\bibfnamefont{W.}~\bibnamefont{Zwerger}},
  \bibinfo{journal}{Rev.~Mod.~Phys.} \textbf{\bibinfo{volume}{80}}, \bibinfo{pages}{885} (\bibinfo{year}{2008}).

\bibitem[{\citenamefont{Greiner et~al.}(2002)\citenamefont{Greiner, Mandel,
  Esslinger, H\"ansch, and Bloch}}]{GreinerBloch2002}
\bibinfo{author}{\bibfnamefont{M.}~\bibnamefont{Greiner \etal}},
  \bibinfo{journal}{Nature} \textbf{\bibinfo{volume}{415}}, \bibinfo{pages}{39}
  (\bibinfo{year}{2002}).

\bibitem[{\citenamefont{Jaksch et~al.}(1998)\citenamefont{Jaksch, Bruder,
  Cirac, Gardiner, and Zoller}}]{JakschZoller1998}
\bibinfo{author}{\bibfnamefont{D.}~\bibnamefont{Jaksch \etal}},
  \bibinfo{journal}{Phys.~ Rev.~ Lett.} \textbf{\bibinfo{volume}{81}},
  \bibinfo{pages}{3108} (\bibinfo{year}{1998}).

\bibitem[{\citenamefont{Hofstetter et~al.}(2002)\citenamefont{Hofstetter,
  Cirac, Zoller, Demler, and Lukin}}]{HofstetterLukin2002}
\bibinfo{author}{\bibfnamefont{W.}~\bibnamefont{Hofstetter \etal}},
  \bibinfo{journal}{Phys.~ Rev.~ Lett.} \textbf{\bibinfo{volume}{89}},
  \bibinfo{pages}{220407} (\bibinfo{year}{2002}).

\bibitem[{\citenamefont{Werner et~al.}(2005)\citenamefont{Werner, Parcollet,
  Georges, and Hassan}}]{WernerHassan2005}
\bibinfo{author}{\bibfnamefont{F.}~\bibnamefont{Werner}},
  \bibinfo{author}{\bibfnamefont{O.}~\bibnamefont{Parcollet}},
  \bibinfo{author}{\bibfnamefont{A.}~\bibnamefont{Georges}}, \bibnamefont{and}
  \bibinfo{author}{\bibfnamefont{S.~R.} \bibnamefont{Hassan}},
  \bibinfo{journal}{Phys. Rev. Lett.} \textbf{\bibinfo{volume}{95}},
  \bibinfo{eid}{056401} (\bibinfo{year}{2005}).

\bibitem[{\citenamefont{Batrouni et~al.}(2002)\citenamefont{Batrouni, Rousseau,
  Scalettar, Rigol, Muramatsu, Denteneer, and Troyer}}]{BatrouniTroyer2002}
\bibinfo{author}{\bibfnamefont{G.~G.} \bibnamefont{Batrouni \etal}},
  \bibinfo{journal}{Phys.~ Rev.~ Lett.} \textbf{\bibinfo{volume}{89}},
  \bibinfo{pages}{117203} (\bibinfo{year}{2002}).

\bibitem[{\citenamefont{Kollath et~al.}(2004)\citenamefont{Kollath,
  Schollw\"ock, von Delft, and Zwerger}}]{KollathZwerger2003}
\bibinfo{author}{\bibfnamefont{C.}~\bibnamefont{Kollath}},
  \bibinfo{author}{\bibfnamefont{U.}~\bibnamefont{Schollw\"ock}},
  \bibinfo{author}{\bibfnamefont{J.}~\bibnamefont{von Delft}},
  \bibnamefont{and} \bibinfo{author}{\bibfnamefont{W.}~\bibnamefont{Zwerger}},
  \bibinfo{journal}{Phys.~ Rev.~ A} \textbf{\bibinfo{volume}{69}},
  \bibinfo{pages}{031601(R)} (\bibinfo{year}{2004}).

\bibitem[{\citenamefont{Rigol et~al.}(2003)\citenamefont{Rigol, Muramatsu,
  Batrouni, and Scalettar}}]{RigolScalettar2003}
\bibinfo{author}{\bibfnamefont{M.}~\bibnamefont{Rigol}},
  \bibinfo{author}{\bibfnamefont{A.}~\bibnamefont{Muramatsu}},
  \bibinfo{author}{\bibfnamefont{G.~G.} \bibnamefont{Batrouni}},
  \bibnamefont{and} \bibinfo{author}{\bibfnamefont{R.~T.}
  \bibnamefont{Scalettar}}, \bibinfo{journal}{Phys. Rev. Lett.}
  \textbf{\bibinfo{volume}{91}}, \bibinfo{pages}{130403}
  (\bibinfo{year}{2003}).

\bibitem[{\citenamefont{Liu et~al.}(2005)\citenamefont{Liu, Drummond, and
  Hu}}]{LiuHu2005}
\bibinfo{author}{\bibfnamefont{X.-J.} \bibnamefont{Liu}},
  \bibinfo{author}{\bibfnamefont{P.~D.} \bibnamefont{Drummond}},
  \bibnamefont{and} \bibinfo{author}{\bibfnamefont{H.}~\bibnamefont{Hu}},
  \bibinfo{journal}{Phys. Rev. Lett.} \textbf{\bibinfo{volume}{94}},
  \bibinfo{eid}{136406} (\bibinfo{year}{2005}).

\bibitem[{\citenamefont{Snoek et~al.}(2008)\citenamefont{Snoek, Titvinidze,
  Toke, Byczuk, and Hofstetter}}]{SnoekHofstetter2008}
\bibinfo{author}{\bibfnamefont{M.}~\bibnamefont{Snoek \etal}},
  \bibinfo{journal}{New J. Phys.} \textbf{\bibinfo{volume}{10}}, 
  \bibinfo{eid}{093008} (\bibinfo{year}{2008}).

\bibitem[{\citenamefont{Helmes et~al.}(2008)\citenamefont{Helmes, Costi, and
  Rosch}}]{HelmesRosch2008}
\bibinfo{author}{\bibfnamefont{R.~W.} \bibnamefont{Helmes}},
  \bibinfo{author}{\bibfnamefont{T.~A.} \bibnamefont{Costi}}, \bibnamefont{and}
  \bibinfo{author}{\bibfnamefont{A.}~\bibnamefont{Rosch}},
  \bibinfo{journal}{Phys. Rev. Lett.} \textbf{\bibinfo{volume}{100}},
  \bibinfo{eid}{056403} (\bibinfo{year}{2008}).

\bibitem[{\citenamefont{Georges et~al.}(1996)\citenamefont{Georges, Kotliar,
  Krauth, and Rozenberg}}]{GeorgesRozenberg1996}
\bibinfo{author}{\bibfnamefont{A.}~\bibnamefont{Georges}},
  \bibinfo{author}{\bibfnamefont{G.}~\bibnamefont{Kotliar}},
  \bibinfo{author}{\bibfnamefont{W.}~\bibnamefont{Krauth}}, \bibnamefont{and}
  \bibinfo{author}{\bibfnamefont{M.~J.} \bibnamefont{Rozenberg}},
  \bibinfo{journal}{Rev. Mod. Phys.} \textbf{\bibinfo{volume}{68}},
  \bibinfo{pages}{13} (\bibinfo{year}{1996}).


\bibitem[{\citenamefont{Scarola et~al.}(2008)\citenamefont{Scarola, Pollet,
  Oitmaa, and Troyer}}]{Scarola08}
\bibinfo{author}{\bibfnamefont{V.~W.}~\bibnamefont{Scarola}}, 
  \bibinfo{author}{\bibfnamefont{L.}~\bibnamefont{Pollet}},
  \bibinfo{author}{\bibfnamefont{J.}~\bibnamefont{Oitmaa}}, \bibnamefont{and}
  \bibinfo{author}{\bibfnamefont{M.}~\bibnamefont{Troyer}},
  \bibinfo{journal}{arXiv:0809.3239}  (\bibinfo{year}{2008}).


\bibitem[{\citenamefont{St\"oferle et~al.}(2006)\citenamefont{St\"oferle,
  Moritz, G\"unter, K\"ohl, and Esslinger}}]{StoeferleEsslinger2006}
\bibinfo{author}{\bibfnamefont{T.}~\bibnamefont{St\"oferle \etal}},
  \bibinfo{journal}{Phys.~ Rev.~ Lett.} \textbf{\bibinfo{volume}{96}},
  \bibinfo{pages}{030401} (\bibinfo{year}{2006}).


\bibitem[{\citenamefont{K\"ohl}(2006)}]{Koehl2006}
\bibinfo{author}{\bibfnamefont{M.}~\bibnamefont{K\"ohl}},
  \bibinfo{journal}{Phys. Rev. A} \textbf{\bibinfo{volume}{73}},
  \bibinfo{eid}{031601(R)} (\bibinfo{year}{2006}).

\bibitem[{\citenamefont{Katzgraber et~al.}(2006)\citenamefont{Katzgraber,
  Esposito, and Troyer}}]{KatzgraberTroyer2006}
\bibinfo{author}{\bibfnamefont{H.~G.} \bibnamefont{Katzgraber}},
  \bibinfo{author}{\bibfnamefont{A.}~\bibnamefont{Esposito}}, \bibnamefont{and}
  \bibinfo{author}{\bibfnamefont{M.}~\bibnamefont{Troyer}},
  \bibinfo{journal}{Phys. Rev. A} \textbf{\bibinfo{volume}{74}},
  \bibinfo{eid}{043602} (\bibinfo{year}{2006}).

\bibitem[{\citenamefont{Dare et~al.}(2007)\citenamefont{Dare, Raymond, Albinet,
  and Tremblay}}]{DareTremblay2007}
\bibinfo{author}{\bibfnamefont{A.-M.} \bibnamefont{Dare}},
  \bibinfo{author}{\bibfnamefont{L.}~\bibnamefont{Raymond}},
  \bibinfo{author}{\bibfnamefont{G.}~\bibnamefont{Albinet}}, \bibnamefont{and}
  \bibinfo{author}{\bibfnamefont{A.-M.~S.} \bibnamefont{Tremblay}},
  \bibinfo{journal}{Phys. Rev. B} \textbf{\bibinfo{volume}{76}},
  \bibinfo{eid}{064402} (\bibinfo{year}{2007}).

\bibitem[{\citenamefont{Mathy and Huse}(2008)}]{MathyHuse2008}
\bibinfo{author}{\bibfnamefont{C.}~\bibnamefont{Mathy}} \bibnamefont{and}
  \bibinfo{author}{\bibfnamefont{D.~A.} \bibnamefont{Huse}},
  \bibinfo{journal}{arXiv:0805.1507}  (\bibinfo{year}{2008}).

\bibitem[{\citenamefont{Koetsier et~al.}(2008)\citenamefont{Koetsier, Duine,
  Bloch, and Stoof}}]{KoetsierStoof2008}
\bibinfo{author}{\bibfnamefont{A.}~\bibnamefont{Koetsier}},
  \bibinfo{author}{\bibfnamefont{R.~A.} \bibnamefont{Duine}},
  \bibinfo{author}{\bibfnamefont{I.}~\bibnamefont{Bloch}}, \bibnamefont{and}
  \bibinfo{author}{\bibfnamefont{H.~T.~C.} \bibnamefont{Stoof}},
  \bibinfo{journal}{Phys. Rev. A} \textbf{\bibinfo{volume}{77}},
  \bibinfo{eid}{023623} (\bibinfo{year}{2008}).

\bibitem[{\citenamefont{{Werner} et~al.}(2006)\citenamefont{{Werner},
  {Comanac}, {de' Medici}, {Troyer}, and {Millis}}}]{WernerMillis2006}
\bibinfo{author}{\bibfnamefont{The
impurity model associated to DMFT is solved using the continuous time quantum
Monte Carlo algorithm P.}~\bibnamefont{{Werner \etal}}},
  \bibinfo{journal}{Phys. Rev. Lett.} \textbf{\bibinfo{volume}{97}},
  \bibinfo{pages}{076405} (\bibinfo{year}{2006}).

\end{thebibliography}

\end{document}